\documentclass[showpacs,preprintnumbers,amsmath,amssymb]{revtex4} 
\usepackage{graphics}
\newcommand{\bm}[1]{\mbox{\boldmath${#1}$}}

\begin{document}
\preprint{APS/123-QED}
\draft
\title{$P$-Wave Charmed-Strange Mesons}
\author{Yukiko Yamada}
\altaffiliation[Present address: ]{Department of Physics, Kyushu University, Fukuoka, Japan.}
\author{Akira Suzuki}
\author{Masashi Kazuyama}
\altaffiliation[Present address: ]{Department of Physics, Nagoya University, Nagoya, Japan.}
\affiliation{%
Department of Physics, Tokyo University of Science, Shinjuku, Tokyo, Japan
}%
\author{Masahiro Kimura}
\affiliation{%
Department of Electronics Engineering, Tokyo University of Science, Suwa, Nagano, Japan
}%
\date{\today}

\begin{abstract}
We examine charmed-strange mesons within the framework of the constituent quark model, focusing on the states with $L=1$. 
We are particularly interested in the mixing of two spin-states that are involved in $D_{s1}(2536)$ and 
the recently discovered $D_{sJ}(2460)$. We assume that these two mesons form a pair of states with $J=1$. 
These spin-states are mixed by a type of the spin-orbit interaction that violates the total-spin conservation. 
Without assuming explicit forms for the interactions as functions of the interquark distance, 
we relate the matrix elements of all relevant spin-dependent interactions to the mixing angle 
and the observed masses of the $L=1$ quartet. 
We find that the spin-spin interaction, among various types of the spin-dependent interactions, plays 
a particularly interesting role in determining the spin structure of $D_{s1}(2536)$ and $D_{sJ}(2460)$.

\vspace{0.5cm}
\end{abstract}
\maketitle

\section{Introduction}

%
%
%
Recently a new charmed-strange meson, $D_{sJ}^{*}(2317)$, has been discovered by the BaBar collaboration 
\cite{aubert03}. 
It was confirmed by the CLEO collaboration \cite{besson03}. 
The CLEO reported another charmed-strange meson called $D_{sJ}(2460)$.   
Finally, these mesons were confirmed by the Belle collaboration \cite{krokovny03,mikami04}. 
The masses and decay properties of $D_{sJ}^{*}(2317)$ and $D_{sJ}(2460)$ have been investigated with particular 
structures assumed for them. 
Essentially there are two types of structures assumed. 
The one is ordinary $q\bar{Q}$ structure and the other an exotic structure such as $KD$ 
molecule \cite{barnes03,lipkin04,mehen04,bicudo05} or tetra-quark configuration 
\cite{silvestre93,terasaki03,chen03,dmitra04}. 
We work with the former structure in this paper. 
Then, these new entries together with $D_{s1}(2536)$ and $D_{s2}(2573)$ which were discovered earlier are expected 
to form a quartet with $L=1$ ($P$-states) of the $c\bar{s}$ (or $s\bar{c}$) system. 
In this expectation, Godfrey studied various properties of $D_{sJ}^{*}(2317)$ and $D_{sJ}(2460)$ 
\cite{godfrey03,godfrey05}, following the works prior to the discoveries of these mesons 
\cite{godfrey85,godfrey91}. 
Also decay modes of $D_{sJ}^{*}(2317)$ and $D_{sJ}(2460)$ were analyzed by Colangero and De Fazio \cite{colangero03}, 
Bardeen {et al.} \cite{bardeen03}, Mehen and Springer \cite{mehen04} and Close and Swanson \cite{close05}. 

With respect to the spin-structure of these mesons, there are four states, ${}^1P_1$, ${}^3P_0$, ${}^3P_1$ 
and ${}^3P_2$, in terms of the $(JLS)$ bases.\footnotemark[2]\footnotetext[2]{We use the ordinary spectroscopic 
notation ${}^{2S+1}L_{J}$ that is used for a two-particle system where $S$, $L$ and $J$ are total spin, 
orbital angular momentum and total angular momentum quantum numbers, respectively.} 
While $D_{sJ}^{*}(2317)$ and $D_{s2}(2573)$ can probably be assigned to ${}^3P_0$ and ${}^3P_2$, respectively, 
$D_{s1}(2536)$ and $D_{sJ}(2460)$ are probably mixtures of ${}^1P_1$ and ${}^3P_1$. 
The extent of the mixing can be parameterized by a mixing angle 
\cite{godfrey03,godfrey05,godfrey85,godfrey91,isgur98,lucha03}. 
In addition to the masses of the mesons, the branching fractions for $B\rightarrow\bar{D}D_{sJ}$ followed 
by the electromagnetic (EM) decays of $D_{sJ}$ have also been measured \cite{krokovny03}. 
The mixing angle are closely related to the EM decay rates of $D_{sJ}$ \cite{godfrey05}.

The purpose of this paper is to examine the spin-structure of the four mesons. 
We use the constituent quark model with the interquark interactions that arise from 
the nonrelativistic expansion of the QCD inspired Fermi-Breit interaction. 
We have five types of interactions in the following sense. 
In addition to the spin-independent interaction that consists of a confining potential 
and the color Coulomb interaction, we have four types of spin-dependent interactions. 
They are the spin-spin, tensor and two types of spin-orbit interactions on which we elaborate in the next paragraph. 
The model is the same as the one used by Godfrey {\it et al.} \cite{godfrey85,godfrey91,isgur98} except that 
we do not assume any explicit forms for the interactions as functions of the distance between the two quarks. 
We treat all spin-dependent interactions perturbatively. 

By the two types of the spin-orbit interactions we mean the ones that are respectively symmetric and 
antisymmetric with respect to the interchange of the two quarks. 
We refer to the former as SLS and the latter as ASLS interactions. 
The SLS interaction commutes with the total spin of the two quarks whereas ASLS interaction does not. 
The ASLS interaction violates the conservation of the total spin. 
This is the agent that induces the mixing of ${}^1P_1$ and ${}^3P_1$. 
The ASLS interaction is proportional to the mass difference between the quarks. 
%
%
%
%
Hence its effect can be substantial when the mass difference is large, 
leading to a specific amount of mixing in the heavy quark limit. 
This is indeed the case with the $c\bar{s}$ (or $s\bar{c}$) system as we will see. 
Historically effects of ASLS interaction were first examined for the $\Lambda-N$ interaction 
and hypernuclei \cite{downs62,londergan71,londergan72}. 
%
%
Regarding the particular role of the spin-orbit interactions in $q\bar{Q}$ systems, we refer to 
a series of works by Schnitzer \cite{schintzer79} and the work by Cahn and Jackson \cite{cahn03} 
in addition to those quoted already \cite{godfrey85,godfrey91,isgur98}. 

As we said above we have five types of interactions. 
On the other hand there are five pieces of experimental data now available, which are the masses of the four mesons 
and the branching ratio of the EM decays. (See Eq. (29).) 
The matrix elements of the five interactions (within the $P$-state sector) can be determined such that 
the five pieces of the experimental data be reproduced. 
At the same time the spin structure of the four mesons can be determined. 
In doing so we do not have to know the radial dependence of the interactions. 
As it turns out the spin-spin interaction, among the four types of the spin-dependent interactions, 
plays a particularly interesting role in relation to the spin structure of $D_{s1}(2536)$ and $D_{sJ}(2460)$. 

We begin Sec. II by defining a nonrelativistic model Hamiltonian which incorporates relativistic corrections 
as various spin-dependent interactions and proceed to determining the matrix elements of the interactions by using 
the mass spectra of the $L=1$ quartet of charmed-strange mesons and the EM decay widths of $D_{sJ}(2460)$. 
In Sec. III we make remarks on the approximations that we use. 
Discussions and a summary are given in the last section.
%
%
In Table I we list the observed charmed-strange mesons that we consider in this paper \cite{pdg04}. 

\begin{widetext}
\begin{table*}
\caption{Summary of observed charmed-strange mesons}
\renewcommand{\arraystretch}{0.8}
\begin{tabular}{cccc}
\hline
Label \quad & \quad Mass (MeV) \quad & \quad Assignment (${}^{2S+1}L_{J}$) \quad & \quad Year of discovery \\
\hline
$D_s^\pm$ \quad        & \quad $1968.3\:\pm\:0.5$ \quad   & \quad ${}^1S_0$ \quad & \quad 1983\quad\cite{chen83}\\
$D_s^{*\pm}$           & \quad $2112.1\:\pm\:0.7$ \quad   & \quad Probably ${}^3S_1$ \quad & \quad 1987\quad\cite{blaylock87} \\
$D_{sJ}^{*}(2317)^\pm$ & \quad $2317.4\:\pm\:0.9$ \quad   & \quad Probably ${}^3P_0$ \quad & \quad 2003\:\quad\cite{krokovny03}\: \\
$D_{sJ}(2460)^\pm$     & \quad $2459.3\:\pm\:1.3$ \quad   & \quad ? \quad & \quad 2003\:\quad\cite{krokovny03}\: \\
$D_{s1}(2536)^\pm$     & \quad $2535.35\:\pm\:0.34$ \quad & \quad ? \quad & \quad 1989\quad\cite{albrecht89} \\
$D_{s2}(2573)^\pm$     & \quad $2572.4\:\pm\:1.5$ \quad   & \quad Probably ${}^3P_2$ \quad & \quad 1994\quad\cite{kubota94} \\
\hline
\end{tabular}
\end{table*}
\end{widetext}

\section{Hamiltonian and Mixing Angle}

We assume that the nonrelativistic scheme is appropriate for the system and relativistic corrections can be treated as first order perturbation. The nonrelativistic expansion of the Fermi-Breit interaction gives us the Hamiltonian for a charmed-strange meson in the form of 
\begin{eqnarray}
\label{hamiltonian}
H&=&H_0+{\bm S}_s\cdot{\bm S}_cV_S(r)+S_{12}V_T(r) \nonumber \\
& & +{\bm L}\cdot{\bm S}V_{LS}^{(+)}(r)+{\bm L}\cdot({\bm S}_s-{\bm S}_c)V_{LS}^{(-)}(r) \quad,
\end{eqnarray}
where ${\bm S}_i$ is the spin operator of the strange quark when $i=s$ and of the charmed quark when $i=c$, 
${\bm S}={\bm S}_s+{\bm S}_c$, $S_{12}$ the tensor operator and ${\bm L}$ the orbital angular momentum operator. 
The lowest-order terms in the nonrelativistic expansion are all in $H_0$ which also contains 
a phenomenological potential to confine the quarks. 
More explicitly $H_0$ reads as  
\begin{equation}
\label{unperturbed_hamiltonian}
H_0=m_s+m_c+\frac{{\bm p}_s^2}{2m_s}+\frac{{\bm p}_c^2}{2m_c}+V_C(r)+V_{conf}(r) \quad,
\end{equation}
where $m_i$ and ${\bm p}_i$ are the mass and momentum of quark $i$, respectively, 
$V_C$ is the color Coulomb interaction and $V_{conf}$ the confinement potential. 
The last two terms of Eq. (\ref{hamiltonian}) are the SLS and ASLS interactions, respectively. 
The spatial functions attached to the operators in Eq. (\ref{hamiltonian}) can be expressed in terms of 
$V_C$ and $V_{conf} $\cite{godfrey91,close80}. 
However, we do not need such explicit expressions of these functions as it will become clear shortly.

We start with the eigenstates of $H_0$ such that
\begin{equation}
\label{shroedinger_equation}
H_0\psi_{nJLS}({\bm r}) = E_{nL}^{(0)}\psi_{nJLS}({\bm r}) \quad,
\end{equation}
where
\begin{equation}
\label{unperturbed_state}
\psi_{nJLS}({\bm r})=R_{nL}(r)\sum_{M=-J}^{J}C_M{\cal Y}_{JLS}^{M}(\theta,\phi) \quad .
\end{equation}
Here $C_M$ are constants such that $\sum_M|C_M|^2=1$ and can be chosen as 
$(2J+1)^{-1/2}$ since there is no preferable direction. 
We concentrate on the $P$-states of $n=1$ with no radial node. 
We denote each of the $L=1$ states with single index $\nu$ according to
\begin{eqnarray}
\label{labeling_of_state}
\nu=\left\{
\begin{array}{l}
1 \\ 2 \\ 3 \\ 4 
\end{array} \right. \qquad \mbox{corresponding to} \qquad \left\{
\begin{array}{l}
{}^1P_1 \\ {}^3P_0 \\ {}^3P_1 \\ {}^3P_2
\end{array} \right. \qquad.
\end{eqnarray}

Next we calculate the matrix elements of $H$ in terms of the bases defined by Eqs. (\ref{shroedinger_equation}) 
and (\ref{unperturbed_state}). 
Nonvanishing matrix elements are
\begin{eqnarray}
\label{matrix_element_of_H}
\begin{array}{l}
H_{11}=M_0-\frac{3}{4}v_S \\
H_{22}=M_0+\frac{1}{4}v_S-2v_{LS}-4v_T \\
H_{33}=M_0+\frac{1}{4}v_S-v_{LS}+2v_T \\
H_{44}=M_0+\frac{1}{4}v_S+v_{LS}-\frac{2}{5}v_T \\
H_{13}=H_{31}=\sqrt{2}\Delta 
\end{array} \qquad,
\end{eqnarray}
where
\begin{eqnarray}
\label{definition_of_M0}
M_0&=&\int d^3r\:\psi_{J1S}^*({\bm r})H_0\psi_{J1S}({\bm r})=E_{1}^{(0)} \\
\label{definition_of_vS}
v_{S}&=&\int_0^{\infty}drr^2V_{S}(r)R_{1}^2(r) \\
\label{definition_of_vLS}
v_{LS}&=&\int_0^{\infty}drr^2V_{LS}^{(+)}(r)R_{1}^2(r) \\
\label{defnition_of_vT}
v_T&=&\int_0^{\infty}drr^2V_T(r)R_{1}^2(r)  \\
\label{definition_of_D}
\Delta &=&\int_0^{\infty}drr^2V_{LS}^{(-)}(r)R_{1}^2(r) \quad. 
\end{eqnarray}
We choose the phases of the wave functions involved in Eq.(\ref{definition_of_D}) such that $\Delta$ is positive.  
Here we have suppressed suffix $n=1$ of the wave functions and the unperturbed $P$-state energy.  
We have ignored the tensor coupling of ${}^3P_2$ state to ${}^3F_2$ state. 
We will remark on this point in the next section. 
Note that the ASLS interaction gives rise to $\Delta\neq 0$ that causes the mixing of $^1P_1$ and $^3P_1$. 

All of the matrix elements of the Hamiltonian that we need are parameterized in terms $M_0$, $v_S$, $v_{LS}$, 
$v_T$ and $\Delta$. 
These five parameters can be determined by the four observed masses and the EM decay rates of $D_{sJ}(2460)$. 
We have no other adjustable parameters. 
In this context we do not need explicit expressions of the radial wave function nor the radial dependence of 
the potential functions.  

The diagonalization of $H$ leads to four states whose masses are given by 
\begin{eqnarray}
\label{M1}
& & {\cal M}_+=\frac{1}{2}\left[2M_0-\frac{1}{2}v_S-v_{LS}+2v_T \right. \nonumber \\
& &  \hspace{10mm} \left.+\left\{\left(v_{LS}-2v_T-v_S\right)^2+8\Delta^2\right\}^{1/2}\;\right] \\[1mm]
\label{M2}
& & {\cal M}_2=M_0+\frac{1}{4}v_S-2v_{LS}-4v_T \\[1mm]
\label{M3}
& & {\cal M}_-=\frac{1}{2}\left[2M_0-\frac{1}{2}v_S-v_{LS}+2v_T \right. \nonumber \\
& &  \hspace{10mm} \left.-\left\{\left(v_{LS}-2v_T-v_S\right)^2+8\Delta^2\right\}^{1/2}\;\right] \\[1mm]
\label{M4}
& & {\cal M}_4=M_0+\frac{1}{4}v_S+v_{LS}-\frac{2}{5}v_T \quad.
\end{eqnarray}
The second and fourth states with ${\cal M}_2$ and ${\cal M}_4$ are pure ${}^3P_0$ and ${}^3P_2$ states, 
respectively. 
We identify them with $D_{sJ}^*(2317)$ and $D_{s2}(2573)$. 
Other two states with ${\cal M}_+$ and ${\cal M}_-$ are composed of ${}^1P_1$ and ${}^3P_1$ states. 
We interpret them as $D_{s1}(2536)$ and $D_{sJ}(2460)$, respectively. 

Let us introduce a mixing angle $\theta$ that represents the extent of the mixing of ${}^1P_1$ and 
${}^3P_1$ states in $D_{s1}(2536)$ and $D_{sJ}(2460)$. 
Following Godfrey and Isgur \cite{godfrey85}, we define $\theta$ by 
\begin{eqnarray}
\label{definition_of_mixing_angle}
\begin{array}{l}
\psi_+({\bm r})=-\psi_{110}({\bm r})\sin\theta+\psi_{111}({\bm r})\cos\theta \\
\psi_-({\bm r})=\psi_{110}({\bm r})\cos\theta+\psi_{111}({\bm r})\sin\theta  
\end{array}
\end{eqnarray}
where $\psi_+$ and $\psi_-$ are the eigenstates that correspond to $D_{s1}(2536)$ and $D_{sJ}(2460)$, 
respectively.
The requirement that the energy eigenvalues for $\psi_\pm$ are ${\cal M}_\pm$ leads to   
\begin{eqnarray}
\label{tan_of_mixing_angle}
\tan(2\theta)=-\frac{2\sqrt{2}\Delta}{v_S-v_{LS}+2v_T} \quad.
\end{eqnarray}
It is understood that $\theta$ lies in the interval of $-\pi/2\le\theta\le0$ 
so that it conforms to the sign convention used in Ref.\cite{godfrey85}. 
Since $-\pi/4\le\theta\le0$ (or $-\pi/2\le\theta\le-\pi/4$) if $(v_S-v_{LS}+2v_T)\ge0$ (or $\le0$), 
we have $\theta\rightarrow0$ (or $\rightarrow-\pi/2$) as $\Delta\rightarrow0$ if $(v_S-v_{LS}+2v_T)\ge0$ (or $\le0$).
In other words, when $(v_S-v_{LS}+2v_T)>0$ (or $<0$), $D_{s1}(2536)$ develops from the ${}^3P_1$ 
(or ${}^1P_1$) state, while $D_{sJ}(2460)$ from the ${}^1P_1$ (or ${}^3P_1$) state due to the ASLS interaction . 

We can express the five parameters, $M_0$, $v_S$, $v_{LS}$, $v_T$ and $\Delta$, in terms of the four observed 
masses and the mixing angle such that 
\begin{eqnarray}
\label{M0}
M_0&=&\frac{1}{4}{\cal M}_{+}+\frac{1}{4}{\cal M}_{-}+\frac{1}{12}{\cal M}_{2}+\frac{5}{12}{\cal M}_{4} \\[1mm]
\label{vS}
v_S&=&-\frac{1}{3}\left(1-2\cos(2\theta)\right){\cal M}_{+}
-\frac{1}{3}\left(1+2\cos(2\theta)\right){\cal M}_{-}  \nonumber \\
& & +\frac{1}{9}{\cal M}_{2}+\frac{5}{9}{\cal M}_{4} \\[1mm]
\label{vLS}
v_{LS}&=&-\frac{1}{8}\left(1+\cos(2\theta)\right){\cal M}_{+}
-\frac{1}{8}\left(1-\cos(2\theta)\right){\cal M}_{-}  \nonumber \\
& & -\frac{1}{6}{\cal M}_{2}+\frac{5}{12}{\cal M}_{4} \\[1mm]
\label{vT}
v_{T}&=&\frac{5}{48}\left(1+\cos(2\theta)\right){\cal M}_{+}
+\frac{5}{48}\left(1-\cos(2\theta)\right){\cal M}_{-}  \nonumber \\
& & -\frac{5}{36}{\cal M}_{2}-\frac{5}{72}{\cal M}_{4} \\[1mm]
\label{delta}
\Delta&=&-\frac{1}{2\sqrt{2}}(M_+-M_-)\sin(2\theta) \quad.
\end{eqnarray}
%
%
Equation (\ref{M0}) states the fact that the mass of the center of gravity 
of the $l=1$ quartet is free from the spin-dependent interactions involved in Eq.(\ref{hamiltonian})
in the lowest order perturbation scheme. 

In order to determine the mixing angle, we consider EM decays of $D_{sJ}(2460)$ 
to $D_s$ and $D_s^*$. 
Generally the $E1$ decay width of a meson composed of quark 1 and anti-quark 2 is given by 
\begin{equation}
\label{decay_formula}
\Gamma(i\rightarrow f+\gamma)=\frac{4e_Q^2}{27}\:k^3(2J_f+1)
\left|\left\langle f\left|r\right|i\right\rangle\right|^2S_{if} \quad,
\end{equation}
where $e_Q$ is the effective charge defined by 
\begin{equation}
\label{effective_charge}
e_Q=\frac{m_1e_2-m_2e_1}{m_1+m_2} \quad,
\end{equation}
$k$ the momentum of the emitted photon 
\begin{equation}
\label{momentum}
k=\frac{M_i^2-M_f^2}{2M_i} \quad,
\end{equation}
and 
\begin{eqnarray}
\label{factor}
S_{if}=\left\{
\begin{array}{ll}
1 \quad & \mbox{for a transition between triplet states,} \\
3 \quad & \mbox{for a transition between singlet states,}
\end{array}\right. 
\end{eqnarray}
is a statistical factor \cite{eichten77}. 
For the decays of $D_{sJ}$, we have 
\begin{eqnarray}
\label{momentum_value}
k=\left\{
\begin{array}{ll}
322.7\:\mbox{MeV} \quad & \quad \mbox{for the decay to }D_s^* \\
442.0\:\mbox{MeV} \quad & \quad \mbox{for the decay to }D_s
\end{array}\right. \quad,
\end{eqnarray}
and $(2J_f+1)S_{if}=3$ for both cases.
%
%
%
Since only the ${}^3P_1$ state in $D_{sJ}$ undergoes the transition to $D_s^*$ and only the ${}^1P_1$ state 
to $D_s$, the matrix element $\left\langle f\left|r\right|i\right\rangle$ is proportional to $\sin\theta$ for 
the decay to $D_s^*$ and to $\cos\theta$ for the decay to $D_s$ \cite{godfrey05}.  
Thus we obtain
\begin{eqnarray}
\label{ratio_to_mixing_angle}
\frac{\Gamma(D_{sJ}\rightarrow D_s^*\gamma)}{\Gamma(D_{sJ}\rightarrow D_s\gamma)}=
\left(\frac{322.7}{442.0}\right)^3\tan^2\theta \quad.
\end{eqnarray}

\begin{figure}
\scalebox{0.471}{\includegraphics*{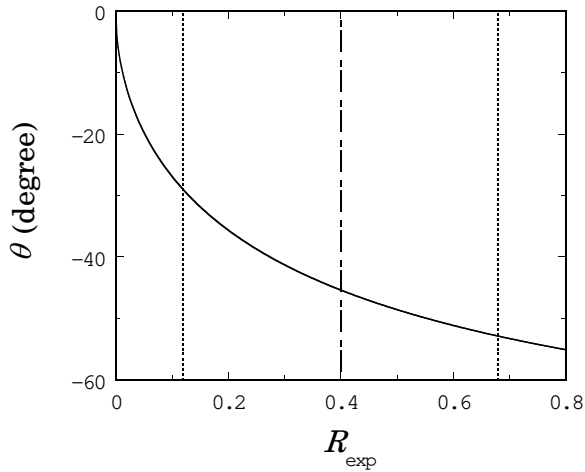}}
\caption{\label{fig_1} Variation of the mixing angle with $R_{\rm exp}$. 
The dot-dashed line shows the value obtained from the central value of $R_{\rm exp}$, 
and the vertical dotted lines indicate the upper and lower values of $R_{\rm exp}$ allowed within 
the statistical errors.}
\end{figure}

%
%
%
The Belle collaboration made the first observation of $B\rightarrow\bar{D}D_{sJ}$ decays and reported 
the branching fractions for $B\rightarrow\bar{D}D_{sJ}$ followed by the EM decays of $D_{sJ}$ \cite{krokovny03}.  
Colangelo {\it et al.} analyzed the data to extract the ratio of branching fractions for the EM decays of 
$D_{sJ}(2460)$ to $D_s$ and $D_s^*$ \cite{colangelo04}. 
They obtained 
\begin{eqnarray}
\label{decay_ratio}
& & R_{\rm exp}\equiv
\left[\frac{\Gamma(D_{sJ}\rightarrow D_s^*\gamma)}{\Gamma(D_{sJ}\rightarrow D_s\gamma)}\right]_{\rm exp}=
0.40\pm0.28 \quad.
\end{eqnarray}
The experimental value has the large statistical errors which results in a large uncertainty in determining 
the mixing angle as can be seen in Fig.~\ref{fig_1}.
The numerical value is 
\begin{eqnarray}
\label{mixing_angle}
\theta=-45.4^\circ\;
\begin{small}
\begin{array}{l}
-7.5^\circ \\[-1mm] +16.4^\circ 
\end{array}
\end{small}
\quad,
\end{eqnarray}
where the upper and lower increments are due to the positive and negative corrections of the statistical errors 
in $R_{\rm exp}$, respectively. 
%
%
%
This may be compared with $-38^\circ$ obtained by Godfrey and Kokoski \cite{godfrey91}, 
and $-54.7^\circ$ that emerges from $\sin\theta=-\sqrt{2/3}$ in the heavy quark limit \cite{isgur98,godfrey05}.  

\begin{figure}
\scalebox{0.471}{\includegraphics*{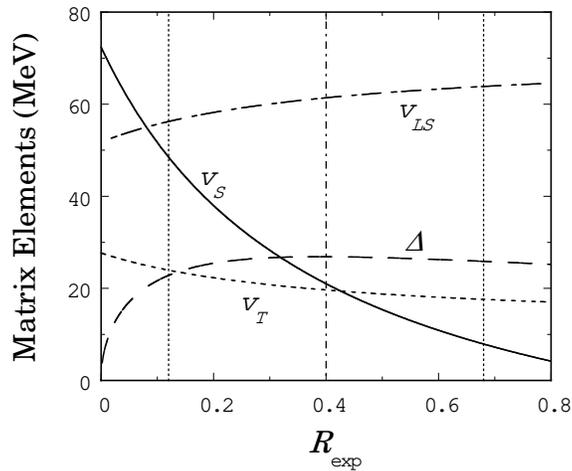}}
\caption{\label{fig_2} The matrix elements calculated by applying the experimental value of 
$R_{\rm exp}$ from Eq.(\ref{M0}) to Eq.(\ref{delta}) with Eq.(\ref{ratio_to_mixing_angle}). 
The dot-dashed line shows the value obtained from the central value of $R_{\rm exp}$, and 
the vertical dotted lines indicate the upper and lower values of $R_{\rm exp}$ allowed 
within the statistical errors.}
\end{figure}

We can calculate $M_0$, $v_S$, $v_{LS}$, $v_T$ and $\Delta$ through Eqs. (\ref{M0}) to (\ref{delta}) by 
fitting the observed masses of Table I and the mixing angle of Eq. (\ref{mixing_angle}). 
Again these quantities are subject to uncertainties due to the statistical errors. 
Using the central values of the observed masses, we obtain 
\begin{eqnarray}
\label{evM0}
& & M_0=2513.6\;\mbox{MeV} \\
\label{evvS}
& & v_S=21.0 \;
\begin{small}
\begin{array}{l}
-13.1 \\[-1mm] +27.5 
\end{array}
\end{small}
\;\mbox{MeV} \\
\label{evLS}
& & v_{LS}=61.4 \;
\begin{small}
\begin{array}{l}
+2.5 \\[-1mm] -5.2 
\end{array}
\end{small}
\;\mbox{MeV} \\
\label{evvT}
& & v_T=19.7 \;
\begin{small}
\begin{array}{l}
-2.1 \\[-1mm] +4.3 
\end{array}
\end{small}
\;\mbox{MeV} \\
\label{evDelta}
& & \Delta=26.9 \;
\begin{small}
\begin{array}{l}
 -1.0 \\[-1mm] -4.1 
\end{array}
\end{small}
\;\mbox{MeV} \quad.
\end{eqnarray}
In Fig.~\ref{fig_2} we show how the matrix elements vary when $R_{\rm exp}$ is varied within the statistical errors. 
Note that the matrix element of the spin-spin interaction is particularly sensitive to the variation of $R_{\rm exp}$. 
Since the sign of $(v_S-v_{LS}+2v_T)$ determines the main spin-states of ${\cal M}_\pm$, it is interesting to see 
the $R_{\rm exp}$ dependence of this quantity shown in Fig. 3. 
\begin{figure}
\scalebox{0.471}{\includegraphics*{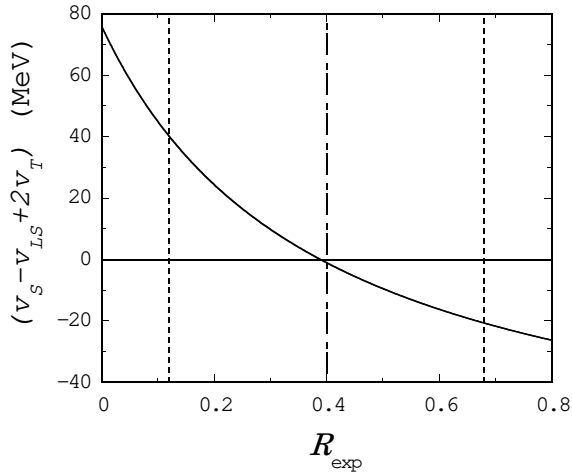}}
\caption{\label{fig_3} The matrix element $(v_S-v_{LS}+2v_T)$ versus $R_{\rm exp}$. 
The value at which the matrix element changes its sign is $0.39$.  
The dot-dashed line shows the value obtained from the central value of $R_{\rm exp}$, and 
the vertical dotted lines indicate 
the upper and lower values of $R_{\rm exp}$ allowed within the statistical errors.}
\end{figure}
We see that $(v_S-v_{LS}+2v_T)$ changes its sign from positive to negative as $R_{\rm exp}$ passes over 0.39. 
If $R_{\rm exp}<0.39$ the main spin-states of $D_{s1}(2536)$ and $D_{sJ}(2460)$ are respectively ${}^3P_1$ and 
${}^1P_1$. 
If $R_{\rm exp}$ exceeds 0.39, these two spin-states are interchanged.

In the nonrelativistic expansion of the Fermi-Breit interaction, the spin-spin interaction contains 
the derivative of the color Coulomb interaction. 
If the color Coulomb interaction is of the form of $1/r$, the spin-spin interaction behaves like 
the delta function near the origin. In that case, the matrix element of the spin-spin interaction will vanish 
in $P$-states because the $P$-state wave functions are strongly suppressed where the interaction acts. 
The real situation, however, is not so simple. 
%
%
%
The singular spin-dependent interactions are smeared out due to the relativistic corrections 
\cite{godfrey85,godfrey91} and the asymptotic freedom.  
The resultant spin-spin interaction will have a well-behaved form at the origin. 
Consequently the matrix element of the spin-spin interaction can become sizable. 
Its magnitude depends on the spatial form of the interaction which in turn depends on how one 
incorporates the relativistic corrections and the asymptotic freedom. 
Equation (\ref{evvS}) that we obtained above is a constraint that the spin-spin interaction has to satisfy.  

Earlier we had experimental information on the effect of the spin-spin interaction on $P$-states of heavy 
quark systems only from the charmonia. 
In first order perturbation theory, we can estimate the matrix element by calculating 
the difference between an weighted average of the masses of ${}^3P$-states and the mass of ${}^1P$-state. 
(See Eq.(\ref{matrix_element_of_H}).) 
For the $c\bar{c}$ system, if we can regard $h_c(1P)$ as the ${}^1P_0$ state \cite{pdg04}, we obtain $-0.85\:$MeV 
for this quantity. 
If one assumes that the spin-spin interaction is inversely proportional to the product of the quark masses 
and that the wave functions of the $c\bar{c}$ system and those of the charmed-strange mesons are the same, 
one obtains about -3 MeV for the charmed-strange mesons. 
The value that emerged from our analysis is much larger in magnitude than this value. 

%
%
%
Let us make remarks on the works of Godfrey and Isgur \cite{godfrey85} and of Godfrey and Kokoski 
\cite{godfrey91} in comparison with the present work.  
They used basically the same Hamiltonian that we used and diagonalized it on the basis of the harmonic oscillator 
eigenstates. 
They assumed explicit forms for the confinement potential and the color Coulomb interaction in terms of which 
the spatial behavior of all spin-dependent interactions can be expressed. 
%
%
%
They accomplished the relativistic corrections by introducing a smearing function which softens 
the singular behavior of the spin-dependent interactions at the origin. 
As a consequence a sizable contribution from the spin-spin interaction to the matrix element for the $P$-state 
emerged.  
They fixed the parameters by fitting observed meson masses known then and predicted unobserved meson masses.  
Although they worked out beyond the perturbation theory, it would be interesting to estimate the matrix elements 
of the spin-dependent interactions perturbatively through Eqs.(\ref{M0}) to (\ref{delta}) from the masses and 
the mixing angle that they obtained for charmed-strange mesons.

The results are summarized in Table II, compared with preceding works by Godfrey and Kokoski \cite{godfrey91} 
and Lucha and Sch\"oberl \cite{lucha03}.   
The last four numbers under the meson symbols in the first row named this work are the observed masses that we used 
to evaluate the matrix elements in our analysis. 
The last four numbers in the other rows are the predicted masses. 
The values in parentheses in the third row are the matrix elements obtained nonperturbatively in 
Ref. \cite{godfrey91}.  
Note that these values are very close to the corresponding ones of the second row, showing that our perturbative 
treatment is adequate.  

The masses of $D_{sJ}^*$ predicted in Refs. \cite{godfrey91} and \cite{lucha03} are much larger than 
the experimental value. 
The matrix elements of the SLS and tensor interactions come into the masses of $D_{sJ}^*$ with negative 
sign as seen in Eq.(\ref{M2}).
In Refs. \cite{godfrey91} and \cite{lucha03} the magnitudes of these matrix elements are very small compared 
with the ones that the experiments require.  
This is why they had approximately $110$ to $120\:$MeV larger masses for $D_{sJ}^*$ compared with the experimental value 
even when one corrects the overestimate of the center of gravity ($M_0$) for the $P$-state masses.  
On the other hand, the matrix elements of the SLS and tensor interactions come with opposite signs for the mass 
of $D_{s2}$. 
This moderates the overestimate of the mass of $D_{s2}$. 
The mass differences between $D_{s1}$ and $D_{sJ}$ in Refs. \cite{godfrey91} and \cite{lucha03} are very small 
as compared with $76\:$MeV of the experimental value. 
This is simply due to the feature that the values of $\Delta$ of Refs. \cite{godfrey91} and \cite{lucha03} are  
much smaller than the one that implied by experiments.  
%

%
%
\begin{widetext}
\begin{table*}
\caption{Matrix elements, the mixing angle $\theta$ and the masses of the $P$-state charmed-strange mesons 
in the columns with the corresponding meson symbols.
They are given in MeV except $\theta$.  
In the first row named this work, the central values of the masses reported by Particle Data Group \cite{pdg04}   
were listed and we used them to obtain the matrix elements.  
The mixing angle was given by Eq.(\ref{mixing_angle}) with the statistical errors suppressed. 
%
%
%
The values of the masses and the mixing angles in the second row are predictions by the 
indicated authors. 
The matrix elements in each row were calculated by substituting these quantities into Eqs.(\ref{M0})-(\ref{delta}).
%
%
%
%
The numbers in the parentheses in the third row are the matrix elements obtained in Ref. \cite{godfrey91}.} 
\renewcommand{\arraystretch}{0.8}
\begin{tabular}{ccccccccccc}
\hline
\quad & \quad $M_0$ \quad & \quad $v_s$ \quad & \quad $v_{LS}$ \quad & \quad $v_T$ \quad & \quad $\Delta$ \quad 
& \quad $\theta$ \quad & \quad $D_{sJ}^*$ \quad & \quad $D_{sJ}$ \quad & \quad $D_{s1}$ \quad & \quad $D_{s2}$\quad\\
\hline
This work \quad      & \quad $2513.6$ \quad   & \quad $21.0$ \quad & \quad $61.4$ \quad & \quad $19.7$ \quad & \quad 
$26.9$ \quad & \quad $-45.4^\circ$ \quad & \quad $2317.4$ \quad & \quad $2459.3$ \quad & \quad $2535.35$ \quad & \quad 
$2572.4$ \quad \\
%
%
Godfrey and Kokoski \cite{godfrey91} & \quad $2563$ \quad   & \quad $13$ \quad & \quad $27$ \quad & \quad  $8$ \quad & \quad 
$3$ \quad & \quad $-38^\circ$ \quad & \quad $2480$ \quad & \quad $2550$ \quad & \quad $2560$ \quad & \quad 
$2590$ \quad \\
                   & \quad $(2564)$ \quad   & \quad $(15)$ \quad & \quad $(27)$ \quad & \quad  $(7)$ \quad & \quad 
$(3)$ \quad & & & & & \\
Lucha and Sch\"oberl \cite{lucha03} & \quad $2531$ \quad   & \quad $14$ \quad & \quad $29$ \quad & \quad  $8$ \quad & \quad 
$4$ \quad & \quad $-44.7^\circ$ \quad & \quad $2446$ \quad & \quad $2515$ \quad & \quad $2517$ \quad & \quad 
$2561$ \quad \\
\hline
\end{tabular}
\end{table*}
\end{widetext}

\section{Validity of the Approximations Used}

We give remarks on the approximations that we have used in Sec.\ II. 
First, we have regarded all spin-dependent interactions as perturbation and obtained their matrix elements as 
given in Eqs.(\ref{evvS})-(\ref{evDelta}). 
A typical mass difference $\Delta M_0$ between two consecutive principal states that emerges from $H_0$ is probably 
like $400\:$MeV to $500\:$MeV. 
The values of $v_S$, $v_{LS}$, $v_T$ and $\Delta$ are much smaller than $\Delta M_0$. 
This justifies our perturbative treatment of the spin-dependent interactions.   

Secondly, we have ignored the tensor coupling of ${}^3P_2$ state to ${}^3F_2$ state. 
The nonvanishing matrix element of the tensor interaction between these states gives rise to an additive correction 
to $H_{44}$ in Eq.(\ref{matrix_element_of_H}) through the second order perturbation. 
Let us have an estimate of the second order correction. 
Note that a quark in a state with $L\ge1$ feels the color Coulomb interaction much less than a quark in an $S$ state. 
This is because the wave function of the former is much suppressed near the origin as compared with the wave function 
of the latter.
Therefore, the $P$- and $F$-state wave functions are not very different from those emerging from the confinement 
potential alone. 
Let us ignore the color Coulomb interaction in obtaining the $P$- and $F$-state wave functions and use the harmonic 
oscillator potential for $V_{conf}$. 
Then the radial parts of nodeless $P$- and $F$-state wave functions are given by 
\begin{eqnarray}
\label{p_state}
R_1(r)&=&\sqrt{\frac{8}{3}}\left[\frac{(\mu\omega)^5}{\pi}\right]^{1/4}r\: 
{\rm e}^{-\mu\omega r^2/2} \\
\label{f_state}
R_3(r)&=&\sqrt{\frac{32}{105}}\left[\frac{(\mu\omega)^9}{\pi}\right]^{1/4}r^3\:
{\rm e}^{-\mu\omega r^2/2} \quad, 
\end{eqnarray} 
where $\omega$ is an oscillator constant and $\mu=(1/m_s+1/m_c)^{-1}$ the reduced mass. 
Remember that $\omega$ is related to the mass difference between two consecutive principal states and, 
hence, $\omega\approx\Delta M_0$. 
Since the tensor interaction can be expressed as  
\begin{equation}
\label{tensor_potential}
V_T(r)=\frac{V_C'(r)-rV_C''(r)}{12m_sm_cr} 
\end{equation}
in terms of the color Coulomb interaction, the needed diagonal and off-diagonal matrix 
elements are given by
\begin{eqnarray}
\label{off_diagonal_element}
& & \left\langle{}^3P_2\left|S_{12}V_T(r)\right|{}^3P_2\right\rangle
=-\frac{8}{45} \sqrt{\frac{(\mu\omega)^3}{\pi}}\frac{\alpha}{m_sm_c} \\
& & \left\langle{}^3P_2\left|S_{12}V_T(r)\right|{}^3F_2\right\rangle
=\frac{16}{15}\sqrt{\frac{6}{35}}\sqrt{\frac{(\mu\omega)^3}{\pi}}\frac{\alpha}{m_sm_c} \quad ,
\end{eqnarray}
where we have used
\begin{equation}
\label{coulomb_potential}
V_C(r)=-\frac{4}{3}\frac{\alpha}{r}
\end{equation}
with the strong coupling constant $\alpha$. 
Thus we obtain 
\begin{eqnarray}
\label{ratio}
\frac{\left\langle{}^3P_2\left|S_{12}V_T(r)\right|{}^3F_2\right\rangle}
{\left\langle{}^3P_2\left|S_{12}V_T(r)\right|{}^3P_2\right\rangle}
=-6\sqrt{\frac{6}{35}}\approx-2.5 \quad.
\end{eqnarray}
If we equate the denominator to the matrix element of the tensor operator times 
the quantity given in Eq.(\ref{evvT}), that is,  
\begin{eqnarray}
\label{diagonal_element_value}
& & \left\langle{}^3P_2\left|S_{12}V_T(r)\right|{}^3P_2\right\rangle
\approx-8\;\mbox{MeV} \quad,
\end{eqnarray}
an approximate magnitude of the off-diagonal element becomes 
\begin{equation}
\left\langle{}^3P_2\left|S_{12}V_T(r)\right|{}^3F_2\right\rangle\approx20\;\mbox{MeV} \quad.
\end{equation}
Since the energy difference between the ${}^3P_2$ and ${}^3F_2$ states is approximately 
$2\times\omega\approx1\;$GeV, the second order correction will be about $0.4\:$MeV, that is, 
about 5\% of the diagonal element for the ${}^3P_2$ state. 
Thus we conclude that the tensor coupling to ${}^3F_2$ state will not change our result appreciably. \\

\section{Discussions and Summary}

We have examined the $P$-state charmed-strange mesons, focusing on the mixing of ${}^1P_1$ and ${}^3P_1$ states 
in $D_{s1}(2536)$ and $D_{sJ}(2460)$ that is caused by the antisymmetric spin-orbit (ASLS) interaction. 
We have treated the spin-dependent interactions that arise from the nonrelativistic expansion of the Fermi-Breit 
interaction perturbatively. 
We have not assumed any explicit forms for the interactions as functions of the interquark distance. 
We have expressed the matrix elements of these interactions in terms of the observed masses of the $P$-state 
quartet and the mixing angle determined from the EM decay rates of $D_{sJ}(2460)$.  

The EM decay rates have large statistical errors. 
If we vary the decay rates within the errors, the mixing angle varies widely. 
The matrix elements of the spin-dependent interactions also vary accordingly. 
The matrix elements of the SLS, tensor and ASLS interactions are relatively stable with the variation of 
the mixing angle, varying only within 20\%. 
On the other hand, the matrix element of the spin-spin interaction varies from $48.5\:$MeV to $7.9\:$MeV 
when the mixing angle varies from one end to the other determined from the EM decay rates with 
the statistical errors. 
%
%
%
Note that Godfrey and Kokoski obtained the matrix element of the spin-spin interaction $15\:$MeV that lies 
in this interval \cite{godfrey91}.  

The matrix element of the spin-spin interaction is particularly sensitive to the mixing angle 
and of crucial importance in determining the dominant states of $D_{s1}(2536)$ and $D_{sJ}(2460)$. 
With the large variation of the mixing angle, the dominant state of $D_{s1}(2536)$ is transferred from 
${}^3P_1$ to ${}^1P_1$ state and that of $D_{sJ}(2460)$ from ${}^1P_1$ to ${}^3P_1$ state. 
%
%
%
This implies that the spin-spin interaction is the most important among the spin-dependent interactions 
for the determination of the dominant states in $D_{s1}(2536)$ and $D_{sJ}(2460)$.  
It will be crucial for their assignments, provided other mechanisms for the mixing such as the coupling to 
decay channels are less significant than what we have discussed \cite{beveren04,simonov04,godfrey04}. 
%

%
%
Our analysis is based on the branching fractions that were obtained from the first observation of 
$B\rightarrow\bar{D}D_{sJ}$ decays by the Belle collaboration.  
The analyses are accompanied by large statistical errors, and so be the mixing angle that is extracted from 
the branching fractions. 
For further discussion of the relationship between the mixing angle and the spin-dependent 
interactions, we need more refined data of the branching fractions by the experimental groups.  

\begin{acknowledgments}
We are grateful to Yuki Nogami for helpful comments on the manuscript. 
A. S. would like to thank Professor T. Udagawa for the warm hospitality extended to him at 
the University of Texas at Austin where the last part of this work was done.   
\end{acknowledgments}

%
%


\begin{references}
\bibitem{aubert03}
B. Aubert {\it et al}., Phys. Rev. Lett. {\bf 90}, 242001 (2003).
\bibitem{besson03}
D. Besson {\it et al}., Phys. Rev. D {\bf 68}, 032002 (2003).
\bibitem{krokovny03}
P. Krokovny {\it et al}., Phys. Rev. Lett. {\bf 91}, 262002 (2003).
\bibitem{mikami04} 
Y. Mikami {\it et al}., Phys. Rev. Lett. {\bf 92}, 012002 (2004).
%
\bibitem{barnes03}
T. Barnes, F. E. Close and H. J. Lipkin, Phys. Rev. D {\bf 68}, 054006 (2003).
\bibitem{lipkin04}
H. J. Lipkin, Phys. Letters {\bf B580}, 50 (2004).
\bibitem{mehen04}
T. Mehen and R. P. Springer, Phys. Rev. D {\bf 70}, 074014 (2004).
\bibitem{close05}
F. E. Close and E. S. Swanson, Phys. Rev. D {\bf 72}, 094004 (2005). 
\bibitem{bicudo05}
P. Bicudo, Nucl. Phys. {\bf A748}, 537 (2005).
\bibitem{silvestre93}
B. Silvestre-Brac and C. Semay, Z. Phys. C {\bf 57}, 273 (1993); C {\bf 59}, 457 (1993). \\ 
C. Semay and B. Silvestre-Brac, Z. Phys. C {\bf 61}, 271 (1994). 
\bibitem{terasaki03}
K. Terasaki, Phys. Rev. D {\bf 68}, 011501(R) (2003). 
\bibitem{chen03}
H. Y. Cheng and W.-S. Hou, Phys. Lett. {\bf B566}, 193 (2003). 
\bibitem{dmitra04}
V. Dmitra$\check{\mbox{s}}$inovi$\acute{\mbox{c}}$, Phys. Rev. D {\bf 70}, 096011 (2004).  
%
\bibitem{godfrey03}
S. Godfrey, Phys. Letters {\bf B568}, 254 (2003).
\bibitem{godfrey05}
%
S. Godfrey, Phys. Rev. D {\bf 72}, 054029 (2005).
%
\bibitem{godfrey85}
S. Godfrey and N. Isgur, Phys. Rev. D {\bf 32}, 189 (1985).
\bibitem{godfrey91}
S. Godfrey and R. Kokoski, Phys. Rev. D {\bf 43}, 1679 (1991).
\bibitem{colangero03} 
P. Colangero and F. De Fazio, Phys. Lett. {\bf B570}, 180 (2003).
\bibitem{bardeen03}
W. A. Bardeen, E. J. Eichten and C. T. Hill, Phys. Rev. D {\bf 68}, 054024 (2003).
%
\bibitem{isgur98}
N. Isgur, Phys. Rev. D {\bf 57}, 4041 (1998).
%
\bibitem{lucha03}
W. Lucha and F. F. Sch\"oberl, Mod. Phys. Lett. {\bf A18}, 2837 (2003).
%
\bibitem{downs62}
B. W. Downs and R. Schrils, Phys. Rev. {\bf 127}, 1388 (1962).
\bibitem{londergan71}
J. T. Londergan and R. H. Dalitz, Phys. Rev. C {\bf 4}, 747 (1971).
\bibitem{londergan72}
J. T. Londergan and R. H. Dalitz, Phys. Rev. C {\bf 6}, 76 (1972).
%
\bibitem{schintzer79}
H. J. Schnitzer, Phys. Lett. {\bf 76B}, 461 (1978); 
Phys. Rev. D {\bf 19}, 1566 (1979); Nucl. Phys. {\bf B207}, 131 (1982).
\bibitem{cahn03}
R. N. Cahn and J. D. Jackson, Phys. Rev. D {\bf 68}, 037502 (2003).
%
\bibitem{pdg04}
S. Eidelman {\it et al.}, Phys. Lett. {\bf B592}, 1 (2004).  
\bibitem{chen83} 
A. Chen {\it et al}., Phys. Rev. Lett. {\bf 51}, 634 (1983).
\bibitem{blaylock87} 
G. T. Blaylock {\it et al}., Phys. Rev. Lett. {\bf 58}, 2171 (1987).
\bibitem{albrecht89} 
H. Albrecht {\it et al}., Phys. Lett. {\bf B230}, 162 (1989).
\bibitem{kubota94} 
Y. Kubota {\it et al}., Phys. Rev. Lett. {\bf 72}, 1972 (1994).
%
\bibitem{close80}
F. E. Close, {\it An Introduction to Quarks and Partons}, Academic Press, New York, (1979).
%
\bibitem{eichten77} 
E. Eichten and K. Gottfried, Phys. Lett. {\bf 66B}, 286 (1977).
\bibitem{colangelo04} 
P. Colangelo, F. De Fazio and R. Ferrandes, Preprint, BARI-TH/04-486 (2004).
\bibitem{beveren04}
E. van Beveren and G. Rupp, Phys. Rev. Lett. {\bf 93}, 202001 (2004).
\bibitem{simonov04}
Yu. A. Simonov and J. A. Tjon, Phys. Rev. D {\bf 70}, 114013 (2004). 
\bibitem{godfrey04}
S. Godfrey, hep-ph/0409236.
\end{references}
\end{document}